\documentclass[conference]{IEEEtran}

\usepackage{array, calc, color, multicol}
\usepackage{algorithm}
\usepackage[noend]{algorithmic} 
\usepackage{graphics, epstopdf, graphicx, epsfig, psfrag, epsf, subfigure}
\usepackage{amssymb, amsfonts, amsmath, times}
\usepackage{multicol,balance, verbatim}
\usepackage{textcomp, mathrsfs, stmaryrd, setspace}
\usepackage{amssymb}

\newcommand{\ie}{{\em i.e., }}

\newcommand{\eg}{{\em e.g., }}
\newcommand{\Eg}{{\em E.g., }}

\newtheorem{theorem}{Theorem}

\newtheorem{example}{Example}

\DeclareGraphicsExtensions{.eps, .pdf, .png, .jpg}   

\newcommand{\Jset}{\mathcal{J}}

\newcommand{\Nset}{\mathcal{N}}
\newcommand{\Hset}{\mathcal{H}}

\IEEEoverridecommandlockouts

\makeatletter
\newcommand{\oset}[2]{%
{\mathop{#2}\limits^{\vbox to -.5\ex@{\kern-\tw@\ex@
\hbox{\scriptsize #1}\vss}}}}
\makeatother

\begin{document}

\title{Device-Centric Cooperation in Mobile Networks}

\author{Hulya Seferoglu, Yuxuan Xing\\
{\small ECE Department, University of Illinois at Chicago}\\
{ \small \tt hulya@uic.edu, yxing7@uic.edu}
}

\maketitle

\begin{abstract}

The increasing popularity of applications such as video streaming in today's mobile devices introduces higher demand for throughput, and puts a strain especially on cellular links. Cooperation among mobile devices by exploiting both cellular and local area connections is a promising approach to meet the increasing demand. In this paper, we consider that a group of cooperative mobile devices, exploiting both cellular and local area links and within proximity of each other, are interested in the same video content. Traditional network control algorithms introduce high overhead and delay in this setup as the network control and cooperation decisions are made in a source-centric manner. Instead, we develop a device-centric stochastic cooperation scheme. Our device-centric scheme; DcC allows mobile devices to make control decisions such as flow control, scheduling, and cooperation without loss of optimality. Thanks to being device-centric, DcC reduces; (i) overhead; \ie the number of control packets that should be transmitted over cellular links, so cellular links are used more efficiently, and (ii) the amount of delay that each packet experiences, which improves quality of service. The simulation results demonstrate the benefits of DcC.

\end{abstract}

\section{\label{sec:intro}Introduction}
The increasing popularity of applications such as video streaming in today's mobile devices introduces higher demand for throughput, and puts a strain especially on cellular links. In fact, cellular traffic is growing exponentially and it is expected to remain so for the foreseeable future \cite{cisco_index}, \cite{ericsson_report}.

Cooperation among mobile devices is a promising approach to meet the increasing throughput demand over cellular links. In particular, when mobile devices are in the close proximity of each other and are interested in the same content, device-to-device connections such as WiFi or Bluetooth can be opportunistically used to construct a cooperative system \cite{microcast}, \cite{microcast_allerton}. Indeed, this scenario is getting increasing interest \cite{microcast}. \Eg a group of friends may be interested in watching the same video on YouTube, or a number of students may participate in an online education class \cite{microcast}. More details about the practicality of this scenario is provided in \cite{microcast}. To better illustrate this setup, we provide the following example.

\begin{figure}[t!]
\centering
\subfigure[No-cooperation]{ \label{fig:intro_example_a} \scalebox{.44}{\includegraphics[bb=0 0 165 180]{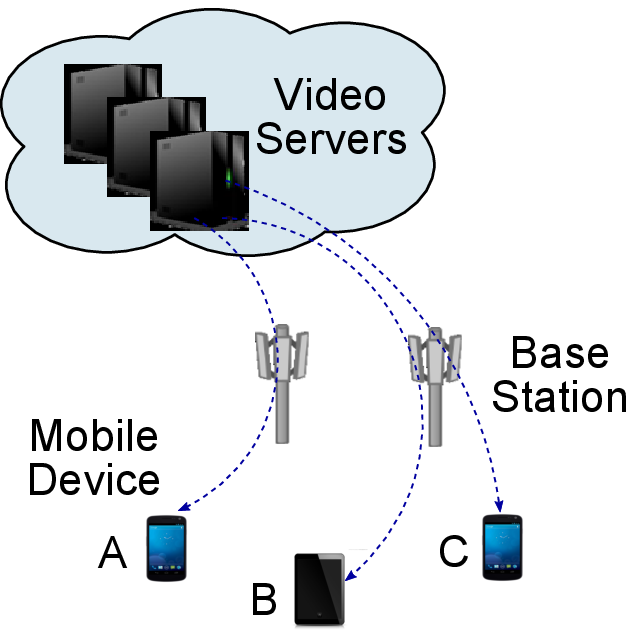}} } \hspace{10pt}
\subfigure[Cooperation]{ \scalebox{.44}{\label{fig:intro_example_b}\includegraphics[bb=0 0 165 180]{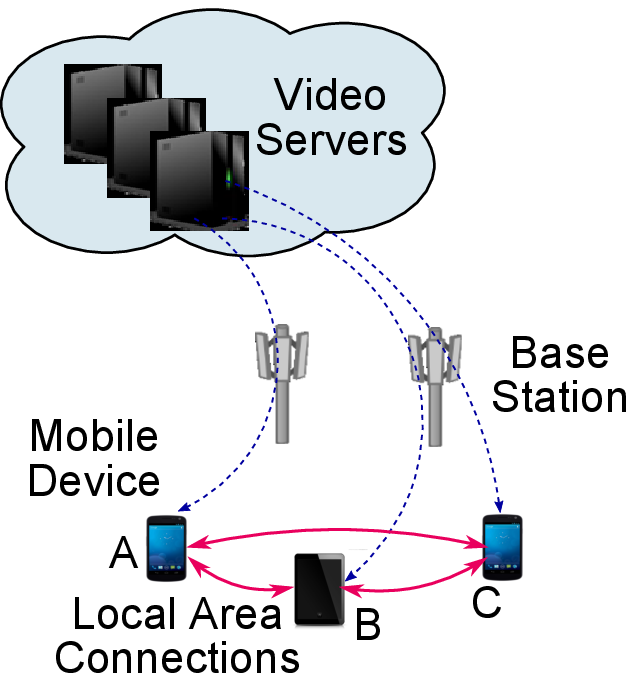}} }
\vspace{-5pt}
\caption{Mobile device users; $A$, $B$, and $C$ are in close proximity, and interested in the same video content. (a) No-cooperation. Each mobile device uses its own cellular link to stream video. (b) Cooperation. Each mobile device uses cellular and local area links simultaneously to stream video.
}
\vspace{-10pt}
\label{fig:intro_example}
\end{figure}

\begin{example}\label{ex1}
Let us consider Fig.~\ref{fig:intro_example}, where mobile device users in close proximity are interested in the same video content. Fig.~\ref{fig:intro_example}(a) shows no-cooperation where each mobile device uses only its cellular link to stream video. For example, if the cellular link rates are 100kbps, each user's streaming rate will be 100kbps. Fig.~\ref{fig:intro_example}(b) shows cooperation, where each mobile device uses cellular and local area links simultaneously (these links operate simultaneously thanks to using different parts of the spectrum) to stream video. Each user downloads 100kbps of video through their cellular connection, and 200kbps from their neighbors. Thus, the streaming rate increases to 300kbps from 100kbps, which is a significant improvement \cite{microcast}, \cite{microcast_allerton}. One important problem, and is the focus of this paper, is the design of a stochastic control algorithm that is efficient in practice in terms of overhead and delay.
\hfill $\Box$
\end{example}

Traditional network control algorithms such as backpressure \cite{tass1}, \cite{tass2}, \cite{neelymoli} make control decisions such as routing and scheduling (and cooperation decision in our problem setup) in a ``source-centric'' manner. In our problem, this corresponds to the case that the servers in the cloud make decisions about (i) the number of video packets that should be pushed to each mobile device, and (ii) the amount of cooperation among mobile devices; \ie the number of packets that each mobile device should transmit to other mobile devices in its neighborhood. In order to make these decisions, video servers should keep track of the states of the mobile devices, which includes queue sizes in mobile devices as well as cellular link qualities towards each mobile device. This puts significant amount of overhead over the cellular links. Furthermore, when there is congestion over the cellular links, the state information, \ie control packets can be delayed significantly, and the video servers may not make timely decisions such as reducing or increasing the rates towards each mobile device. This increases end-to-end delay, which may not fulfill quality of service (QoS) requirements of video streaming applications.

In this paper, we develop a device-centric cooperation scheme to determine the number of video packets each mobile device should receive via cellular links as well as from its neighbors.
Our approach is grounded on a network utility maximization (NUM) formulation of the problem and its solution \cite{tutorial_doyle}. The solution decomposes into several parts with an intuitive interpretation, such as flow control, scheduling over cellular links, and cooperation and scheduling over local area links. Based on the structure of the decomposed solution, we develop a stochastic algorithm; Device-Centric Cooperation; DcC.
The following are the key contributions of this work:
\begin{itemize}
  \item We consider a scenario where a group of cooperative mobile devices, exploiting both cellular and local area links, are within proximity of each other, and are interested in the same content. We propose a novel ``device-centric cooperation'' scheme for this scenario.
  \item We develop network utility maximization (NUM) formulation of the device-centric problem, and provide its decomposed solution. Based on the structure of the decomposed solution, we develop a stochastic device-centric algorithm; DcC. We show that DcC moves the functionality required for cooperation to mobile devices without loss of optimality.
  \item We evaluate our scheme via simulations for multiple mobile devices. The simulation results confirm that DcC reduces; (i) overhead; \ie the number of control packets that should be transmitted over cellular links, and (ii) the amount of delay that each packet experiences.
\end{itemize}

The structure of the rest of the paper is as follows. Section~\ref{sec:system} gives an overview of the system model. Section~\ref{sec:NUM} presents the NUM formulation of our device-centric scheme. Section~\ref{sec:DcC} presents the stochastic device-centric cooperation algorithm; DcC. Section~\ref{sec:DcC_vs_ScC} evaluates DcC. Section~\ref{sec:related} presents related work. Section~\ref{sec:conclusion} concludes the paper.

\section{\label{sec:system}System Model}
In this section, we provide an overview of the device- and source-centric cooperation models demonstrated in Fig.~\ref{fig:forward_backward-system}.\footnote{Note that we provide the source-centric model in addition to our device-centric model so that we can make a connection and comparison between device- and source-centric schemes in the rest of the paper.} First, we provide a cooperative system setup that are common to both device- and source-centric models.

\subsection{Cooperative System}
\begin{figure}[t!]
\centering
\subfigure[Cooperative System]{ \label{fig:intro_example_c} \scalebox{.46}{\includegraphics[bb=0 0 165 180]{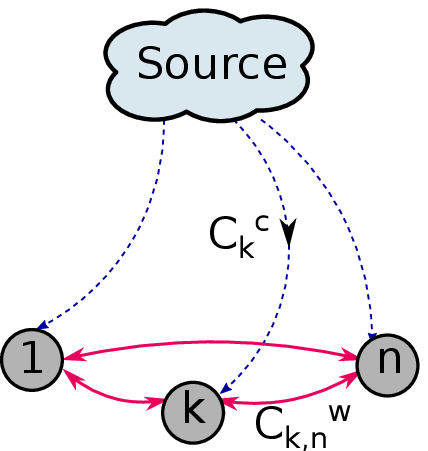}} } 
\subfigure[Source-Centric]{ \label{fig:intro_example_a} \scalebox{.46}{\includegraphics[bb=0 0 165 180]{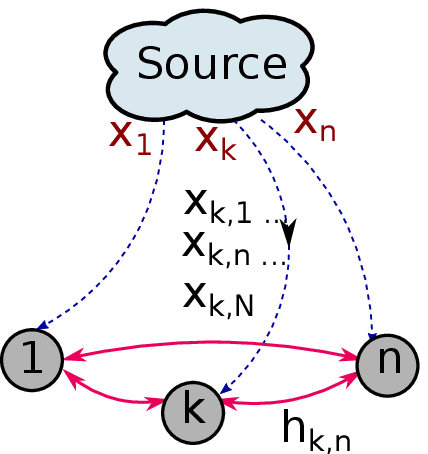}} } 
\subfigure[Device-Centric]{ \scalebox{.42}{\label{fig:intro_example_b}\includegraphics[bb=0 0 165 180]{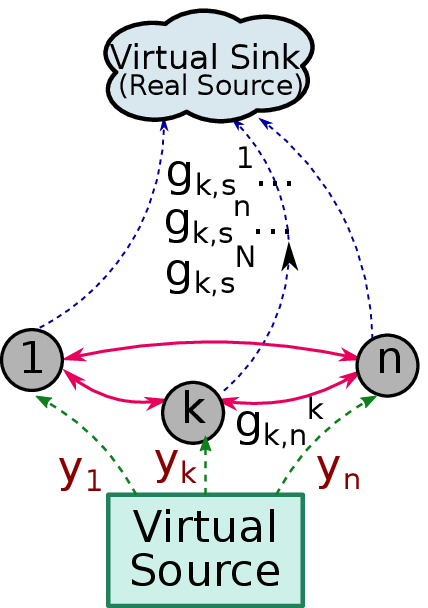}} }
\vspace{-5pt}
\caption{(a) Cooperative system. (b) Source-centric cooperation. (c) Device-centric cooperation.}
\vspace{-10pt}
\label{fig:forward_backward-system}
\end{figure}

{\em Setup:} We consider a cooperative system shown in Fig.~\ref{fig:forward_backward-system}(a), where each mobile device is able to connect to the Internet via cellular links\footnote{Note that our device-centric scheme is generic enough to include Internet connections via WiFi, but we only focus on cellular links for Internet connection in this paper to make the presentation and analysis simple.}, and forward packets to other mobile devices through the local links, \eg Bluetooth or WiFi.

The cooperative system consist of $N$ mobile devices and a source node. Note that the source node represents video servers, proxies, and base stations. This representation allows us to focus on the bottlenecks of the system, namely cellular links from the base station to the mobile devices and the local area links \cite{microcast_allerton}. $\Nset$ is the set of the mobile devices, where $N = |\Nset|$. The mobile devices are interested in the same content and they construct a cooperating group.\footnote{We consider that all mobile devices volunteer to cooperate without any malicious activity. This is possible in our setup due to existing social ties as the mobile device users are in close proximity to each other.}
We consider that time is slotted and $t$ refers to the beginning of slot $t$.

{\em Cellular Links:} Each mobile device $k \in \Nset$ is connected to the Internet via its cellular link. At slot $t$, $\boldsymbol C^{c}(t)$ is the channel state vector of the cellular links, where $\boldsymbol C^{c}(t) = \{C_{1}^{c}(t), ..., C_{k}^{c}(t), ..., C_{N}^{c}(t)\}$. We assume that $C_{k}^{c}(t)$ is the state of the cellular links to mobile node $k$. We consider that cellular links towards different mobile devices are interference free as interference could be handled by base stations. Let $\Gamma_{\boldsymbol C^{c}(t)}$ denote the set of the link transmission rates feasible at time slot $t$ for channel state $\boldsymbol C^{c}(t)$.

{\em Local Area Links:}
In our setup, we consider that mobile devices are in close proximity and they hear each other. Therefore, in the local area, each mobile device can connect to another device directly. This gives us a fully connected topology. Depending on the underlying technology, local area transmissions can be unicast (\eg Bluetooth, or WiFi) or broadcast (can be achieved by extending WiFi \cite{microcast}). In our formulations, we consider both unicast and broadcast transmissions in the local area. We consider protocol model in our formulations \cite{gupta_interference_model}, where each mobile device can either transmit or receive at the same time. Since our local area network is fully connected, only one mobile device can transmit in a slot.

At slot $t$, $\boldsymbol C^{w}(t)$ is the channel state vector of the local area links, where $\boldsymbol C^{w}(t) = \{C_{1,2}^{w}(t), ..., C_{k,n}^{w}(t), ..., C_{N-1,N}^{w}(t)\}$. We assume that $C_{k,n}^{w}(t)$ is the state of the wireless link between node $k$ and $n$. Let $\Gamma_{\boldsymbol C_{u}^{w}(t)}$ denote the set of the link transmission rates feasible at time slot $t$ for channel state $\boldsymbol C^{w}(t)$ for unicast transmission. Similarly, $\Gamma_{\boldsymbol C_{b}^{w}(t)}$ denote the set of the link transmission rates feasible at time slot $t$ for channel state $\boldsymbol C^{w}(t)$ for broadcast transmission.

\subsection{Source-Centric Model}
The source-centric cooperation model is shown in Fig.~\ref{fig:forward_backward-system}(b), where the source node transmits a video flow to a set of mobile devices $\Nset$. The flow generation rate at the source for mobile device $k$ is $x_{k}(t)$, $k \in \Nset$. $x_{k}(t)$ is i.i.d. over the slots and their expected values; $A_{k} = E[x_{k}(t)]$, $E[x_{k}(t)^{2}]$ are finite. Note that even if all mobile devices are interested in the same content, they may receive the content at different rates. In video streaming applications, this corresponds to different levels of video quality. Flow rate $x_k(t)$ is associated with a utility function $U_k(x_k(t))$, which we assume to be strictly concave function of $x_k(t)$.

Flow rate over the cellular link towards node $k$ is $\max_{n \in \Nset}\{x_{k,n}(t)\}$, where $x_{k,k}(t)$ is the rate towards node $k$ to help node $k$, while $x_{k,n}(t)$, $k\neq n$ is the rate towards node $k$ to help node $n$. The flow rate over the cellular link is maximum of the rates, \ie $\max_{n \in \Nset}\{x_{k,n}(t)\}$  as all mobile devices are interested in the same content. Note that $x_{k,k}(t)$ is the rate over the cellular link towards node $k$, while $x_k(t)$ is the flow generation rate for device $k$. Flow rate over the local area link from node $k$ to node $n$ is $h_{k,n}(t)$, $k\neq n$. Note that $h_{k,n}(t)$ is to help node $n$ using node $k$ as a relay.

In the source-centric model, at time slot $t$, queue $\mu_{k}(t)$ is constructed at the source, and it queues packets that will be transmitted to node $k$, and changes according to following dynamics at every time slot $t$.
\begin{align} \label{eq:real_queue_evolutionScC1}
\mu_{k}(t+1) \leq \max [\mu_{k}(t) - \sum_{n \in \Nset}{x}_{n,k}(t), 0] + {x}_{k}(t)
\end{align}
At time slot $t$, queue $\nu_{n,k}(t)$ is the queue size at mobile device $n$, and it queues the packets that should be transmitted to node $k$. $\nu_{n,k}(t)$ changes according to following dynamics at every time slot $t$.
\begin{align} \label{eq:real_queue_evolutionScC2}
\nu_{n,k}(t+1) \leq \max [\nu_{n,k}(t) - {h}_{n,k}(t), 0] + {x}_{n,k}(t)
\end{align}

\subsection{Device-Centric Model}
In the device-centric model shown in Fig.~\ref{fig:forward_backward-system}(c), a virtual source is added to the system and the real source becomes a virtual sink. Node $k$ receives packets with rate $y_k(t)$ from the virtual source and forwards these packets to the virtual sink and other mobile devices. The transmission rate over the cellular link from node $k$ to the virtual sink is $\max_{n \in \Nset}\{g_{k,s}^{s}(t)\}$. The transmission rate from node $k$ to $n$ is $g_{k,n}^{k}(t)$.

Note that the flow rates; $y_k(t)$, $g_{k,s}^{n}(t)$, $g_{k,n}^{k}(t)$ are virtual flow rates. In our device-centric scheme, these virtual flow rates are used to determine the real flow values; $x_k(t)$, $x_{k,n}(t)$, $h_{k,n}(t)$ as explained in Section~\ref{sec:DcC}.

In the device-centric model, at time slot $t$, queue $\lambda_{k}(t)$ is a virtual queue size constructed at node $k$. $\lambda_{k}(t)$ changes according to following dynamics at every time slot $t$.
\begin{align} \label{eq:lambda_queue_evolution}
& {\lambda}_{k}(t+1) \leq  \max [{\lambda}_{k}(t) - {g}_{k,s}^{k}(t) - \sum_{n \in \Nset - \{k\}} {g}_{k,n}^{k}(t), 0] + \nonumber \\
& {y}_{k}(t)
\end{align} At time slot $t$, queue $\eta_{n,k}(t)$ is a virtual queue size constructed at node $n$. $\eta_{n,k}(t)$ changes according to following dynamics at every time slot $t$.
\begin{align} \label{eq:eta_queue_evolution}
{\eta}_{n,k}(t+1) \leq \max [{\eta}_{n,k}(t) - {g}_{n,s}^{k}(t), 0] + {g}_{k,n}^{k}(t)
\end{align} In addition to the virtual queues $\lambda_{k}(t)$ and $\eta_{n,k}(t)$, a real queue $Q_{n,k}(t)$ is constructed at node $n$ and evolves according to the following dynamics at every time slot $t$.
\begin{align} \label{eq:real_queue_evolution}
Q_{n,k}(t+1) \leq \max [Q_{n,k}(t) - {h}_{n,k}(t), 0] + {x}_{n,k}(t)
\end{align} Note that ${h}_{n,k}(t)$ is the amount of the real outgoing traffic from node $n$ to $k$ (\ie from queue $Q_{n,k}$), and ${x}_{n,k}(t)$ is the amount of the real incoming traffic to node $n$ from the source (\ie to the queue $Q_{n,k}$). The relationship between the real and virtual queues as well as real and virtual flows are provided in Section~\ref{sec:DcC}.

\section{\label{sec:NUM} Device-Centric NUM}
In this section, we formulate the device-centric network utility maximization (NUM) framework. This approach sheds light into the structure of the our stochastic algorithm DcC, which we present in the next section.\footnote{Note that NUM optimizes the average values of the parameters that are defined in Section~\ref{sec:system}. By abuse of notation, we use a variable, \eg $\phi$ as the average value of $\phi(t)$  in our NUM formulation if both $\phi$ and $\phi(t)$ refers to the same parameter.}

\subsection{\label{sec:NUM_Formulation} Formulation}
We provide NUM formulations for (i) unicast and (ii) broadcast transmissions in the local area. For unicast setup, the NUM formulation is P-Unicast:
\begin{align} \label{opt:eq1}
\max_{\boldsymbol y, \boldsymbol g} \mbox{ } & \sum_{k \in \Nset} U_{k}(y_k)  \nonumber \\
\mbox{s.t.} \mbox{ }  & g_{k,s}^{k} + \sum_{n \in \Nset-\{k\}} g_{k,n}^{k} = y_{k}, \mbox{ } \forall k \in \Nset \nonumber \\
& g_{n,s}^{k} = g_{k,n}^{k}, \mbox{ } \forall k \in \Nset, n \in \Nset-\{k\} \nonumber \\
& \{\max_{n \in \Nset} \{g_{k,s}^{n}\}\}_{\forall k \in \Nset} \in \Gamma_{C^c},  \nonumber \\
& \{g_{k,n}^{k}\}_{\forall k \in \Nset, n \in \Nset -\{k\}} \in \Gamma_{C_{u}^{w}}.
\end{align}
The objective of P-Unicast is to determine $\boldsymbol y$ $= \{y_k\}_{k \in \Nset}$, $\boldsymbol g$ $= \{g_{n,s}^{k}\}_{k \in \Nset, n \in \Nset}$ which maximize the total utility function; $\sum_{k \in \Nset} U_{k}(y_k)$.
The first constraint is the flow conservation constraint at node $k$; $y_k$ is the incoming traffic rate from virtual source to node $k$, and $g_{k,s}^{k} + \sum_{n \in \Nset-\{k\}} g_{k,n}^{k}$ is the outgoing traffic rate from node $k$ to the virtual sink and the neighbors. The second constraint is the flow conservation constraint at node $n$ for node $k$'s flow; $g_{k,n}^{k}$ is the incoming flow rate to node $n$ from node $k$, and $g_{n,s}^{k}$ is the flow rate from node $n$ towards virtual sink. The last two constraints are the capacity constraints over cellular and local links.

For broadcast setup, the NUM formulation is P-Broadcast. The objective function and the first three constraints of P-Broadcast is the same as P-Unicast in Eq.~(\ref{opt:eq1}). The rest of the constraints of P-Broadcast:
\begin{align} \label{opt:eq2}
&g_{k,n}^{k} \leq \sum_{\Jset \in \Hset | k \in \Jset, n \notin \Jset} f_{n,\Jset},  \mbox{ } \forall k \in \Nset, n \in \Nset - \{k\} \nonumber \\
&\{f_{n, \Jset}\}_{\forall n \in \Nset, \Jset \in \Hset | n \notin \Jset} \in \Gamma_{C_{b}^{w}}.
\end{align}
The first constraint in Eq.~(\ref{opt:eq2}) relates the broadcast transmission rate to the link rate. Let $\Jset$ be a set of nodes, and $\Hset$ be the set of node combinations, \ie $\Jset \in \Hset$. If packets are broadcast from node $n$ to node set $\Jset$, each node $k \in \Jset$ can receive the packets (depending on the loss probability). In the device-centric system, this corresponds to simultaneous transmission from nodes in $\Jset$ to node $k$. $f_{n,\Jset}$  is the broadcast rate in the source-centric system. Since there may be different $\Jset$ sets which contain node $k$, $f_{n,\Jset}$ is summed $\forall \Jset \in \Hset | k \in \Jset, n \notin \Jset$ to determine $g_{k,n}^{k}$. The second constraint in Eq.~(\ref{opt:eq2}) is the broadcast capacity constraint.

\subsection{\label{sec:NUM_Solution} Solution}
Lagrangian relaxation of the first two constraints of both Eq.~(\ref{opt:eq1}) and Eq.~(\ref{opt:eq2}) gives the following Lagrange function:
\begin{align} \label{relax:eq1}
L & = \sum_{k \in \Nset} U_{k}(y_{k}) + \sum_{k \in \Nset} \lambda_{k} (g_{k,s}^{k} + \sum_{n \in \Nset - \{k\}} g_{k,n}^{k} - y_k) + \nonumber \\
& \sum_{k \in \Nset} \sum_{n \in \Nset -\{k\}} \eta_{n,k}(g_{n,s}^{k} - g_{k,n}^{k})
\end{align} where $\lambda_{k}$ and $\eta_{n,k}$ are the Lagrange multipliers. Note that $\lambda_{k}$ and $\eta_{n,k}$ represent the virtual queue sizes defined by Eqs.~(\ref{eq:lambda_queue_evolution}),(\ref{eq:eta_queue_evolution}). The values of $\lambda_{k}$ and $\eta_{n,k}$ are tracked at nodes $k$ and $n$, respectively. Note that these values are virtual values, and a counter is sufficient to keep track of these values.

Eq.~(\ref{relax:eq1}) can be decomposed into several intuitive sub-problems such as rate control, and scheduling. First, we solve the Lagrangian function with respect to $y_k$:
\begin{align} \label{eq:sol_1}
y_k = (U_{k}')^{-1}(\lambda_{k})
\end{align} where $(U_{k}')^{-1}$ is the inverse of the derivative of  $U_k$. Since $U_k$ is strictly concave function of $y_k$, $y_k$ is inversely proportional to $\lambda_{k}$. This means that when the queue size $\lambda_{k}$ increases, $y_k$ should reduce.
In the system implementation, node $k$ requests $y_k$ packets from the real source (\eg video server).

Second, we solve the Lagrangian for $g_{k,s}^{k}$ and $g_{n,s}^{k}$:
\begin{align} \label{eq:sol_3}
\max_{\boldsymbol g} \mbox{ } & \sum_{k \in \Nset} [ \lambda_{k} g_{k,s}^{k} +  \sum_{n \in \Nset-\{k\}} \eta_{k,n}g_{k,s}^{n}]  \nonumber \\
\mbox{s.t.} \mbox{ }  & \{\max_{n \in \Nset} \{g_{k,s}^{n}\}\}_{\forall k \in \Nset} \in \Gamma_{C^c}, \mbox{ }
\end{align} After $g_{k,s}^{k} $ and $g_{k,s}^{n}$ are determined, node $k$ requests $\max_{n \in \Nset}\{g_{k,s}^{n}\}$ packets from the source through its cellular link. Note that $g_{k,s}^{k} $ and $g_{k,s}^{n}$ are different from $y_k$ as $y_k$ is the total flow rate requested by node $k$ and this rate can be transmitted through both its cellular link or from the neighboring nodes, while $g_{k,s}^{k} $ and $g_{k,s}^{n}$ are the rates over cellular links.

Finally, we solve the Lagrangian with respect to $g_{k,n}^{k}$. Note that the solutions in Eq.~(\ref{eq:sol_1}) and Eq.~(\ref{eq:sol_3}) holds for both P-Unicast and P-Broadcast. However, the solutions of P-Unicast and P-Broadcast with respect to $g_{k,n}^{k}$ differ as explained next. The solution of P-Unicast with respect to $g_{k,n}^{k}$ is: $\max_{\boldsymbol g} \mbox{ } \sum_{k \in \Nset} \sum_{n \in \Nset-\{k\}} (\lambda_{k} - \eta_{n,k})g_{k,n}^{k}$ subject to the last two constraints of Eq.~(\ref{opt:eq1}). The solution of P-Broadcast with respect to $g_{k,n}^{k}$ is: $\max_{\boldsymbol g} \mbox{ } \sum_{k \in \Nset} \sum_{n \in \Nset-\{k\}} (\lambda_{k} - \eta_{n,k}) g_{k,n}^{k}$ subject to all the constraints in Eq.~(\ref{opt:eq2}).

Next, we design our stochastic algorithm; Device-Centric Cooperation (DcC) based on the structure of the decomposed NUM solutions, \ie Eq.~(\ref{eq:sol_1}),(\ref{eq:sol_3}) as well as the local area scheduling solution presented above.

\section{\label{sec:DcC} Device-Centric Cooperation (DcC)}
Now, we provide our Device-Centric Cooperation (DcC) algorithm  which includes {\em rate control}, {\em cellular link scheduler} and {\em cooperation \& local area link scheduler}. Note that both unicast and broadcast setups have the same rate control and cellular link scheduling parts. The only different part is the cooperation \& local area link scheduling as explained later.

\underline{Device-Centric Cooperation (DcC):}
\begin{itemize}
 \item {\em Rate Control:} At every time slot $t$, the rate controller at node $k$ determines the number of packets that should be requested from the source according to;
\begin{align} \label{eq:rate_control}
\max_{\boldsymbol {{y}}} & [MU_{k}({y}_k(t)) -  {\lambda}_{k}(t) {y}_{k}(t) ] \nonumber \\
\mbox{s.t. } &  {y}_k(t) \leq R_{k}^{max}
\end{align} where $R_{k}^{max}$ is be a positive constant larger than the cellular rate from the actual source, and $M$ is a large positive constant. The values of $R_{k}^{max}$ and $M$ are important for the stability of the DcC algorithm \cite{thisTechRep}.
${y}_k(t)$ is the number of packets that will be requested from the source.

 \item {\em Cellular Link Scheduler:} At every time slot $t$, the cellular link scheduler at node $k$  determines the number of packets requested through the cellular links.
\begin{align} \label{eq:cellular_scheduling}
\max_{\boldsymbol {{g}}} & \mbox{  }  {\lambda}_{k}(t) {g}_{k,s}^{k}(t) + \sum_{n \in \Nset-\{k\}} ({\eta}_{k,n}(t) - Q_{k,n}(t) ){g}_{k,s}^{n}(t) \nonumber \\
  \mbox{s.t. } & \{{g}_{k,s}^{n}(t)\}_{\forall n \in \Nset} \in \Gamma_{\boldsymbol C^c(t)}.
  \end{align} After $g_{k,s}^{k}(t)$ and ${g}_{k,s}^{n}(t)$ are determined, the real flow rates are determined as ${x}_{k,k}(t) = {g}_{k,s}^{k}(t)$ and ${x}_{k,n}(t) = {g}_{k,s}^{n}(t) - \beta$, where $\beta > 0$ can be chosen to be arbitrarily small, and $\max_{n \in \Nset}\{x_{k,n}(t)\}$ amount of video packets are requested from the source by node $k$.

 \item {\em Cooperation \& Local-Area Link Scheduler for Unicast:} At time slot $t$, the link rate ${g}_{k,n}^{k}(t)$ is determined by;
\begin{align} \label{eq:local_area_scheduling_unicast}
 \max_{\boldsymbol {{g}}} & \mbox{  }  \sum_{k \in \Nset} \sum_{n \in \Nset-\{k\}} [{\lambda}_{k}(t) - {\eta}_{n,k}(t) + Q_{n,k}(t) ]  {g}_{k,n}^{k}(t)   \nonumber \\
\mbox{s.t.} \mbox{ }  & \{{g}_{k,n}^{k}(t)\}_{\forall k \in \Nset, n \in \Nset -\{k\}} \in \Gamma_{\boldsymbol C_{u}^{w}(t)}.
\end{align} After ${g}_{k,n}^{k}(t)$ is determined, ${h}_{n,k}(t) = {g}_{k,n}^{k}(t)$ amount of video packets is requested from node $n$ by node $k$.

 \item {\em Cooperation \& Local-Area Link Scheduler for Broadcast:}
At time slot $t$, the link broadcast rate is determined by;
\begin{align} \label{eq:local_area_scheduling_broadcast}
 \max_{\boldsymbol {{f}}} & \mbox{  }  \sum_{k \in \Nset} \sum_{n \in \Nset-\{k\}} \sum_{\Jset \in \Hset | k \in \Jset, n \notin \Jset}  [{\lambda}_{k}(t) - {\eta}_{n,k}(t) + \nonumber \\
& Q_{n,k}(t) ] {f}_{n,J}(t)   \nonumber \\
\mbox{s.t.} \mbox{ }  & \{{f}_{n,\Jset}(t)\}_{\forall n \in \Nset, \Jset \in \Hset | k \notin \Jset} \in \Gamma_{\boldsymbol C_{b}^{w}(t)}
\end{align}
After ${f}_{n,\Jset}(t)$ is determined, ${f}_{n,\Jset}(t)$ amount of video packets are transmitted from node $n$ to nodes in $\Jset$. The optimum value of ${g}_{k,n}^{k}(t)$ is ${g}_{k,n}^{k}(t) = \sum_{\Jset \in \Hset | k \in \Jset, n \notin \Jset} f_{n,\Jset}(t)$, $\forall k \in \Nset, n \in \Nset - \{k\} $. Therefore, the real transmission rate of over each link is equal to
${h}_{n,k}(t) = {g}_{k,n}^{k}(t) = \sum_{\Jset \in \Hset | k \in \Jset, n \notin \Jset} f_{n,\Jset}(t)$, $\forall k \in \Nset, n \in \Nset - \{k\} $.

\end{itemize}

\begin{theorem}\label{eec_theorem1}
If channel states are i.i.d. over time slots, and the arrival rates $E[y_{t}(t)] = A_k, \forall k \in \Nset$ are interior of the stability region of cellular and local area links, then DcC stabilizes the network and the total average queue sizes, including both virtual and real queues, are bounded for both unicast and broadcast setups.
\end{theorem}
{\em Proof:} The proof is provided in \cite{thisTechRep}.
$\blacksquare$

\begin{theorem}\label{eec_theorem2}
If the channel states are i.i.d. over time slots, and the traffic arrival rates are controlled by the rate control algorithm in Eq.~(\ref{eq:rate_control}), then the admitted flow rates converge to the utility optimal operating point with increasing $M$.
\end{theorem}
{\em Proof:} The proof is provided in \cite{thisTechRep}.
$\blacksquare$

\section{\label{sec:DcC_vs_ScC} Evaluation of Device-Centric Cooperation}
In this section, we evaluate our DcC algorithm as compared to Source-Centric Cooperation (ScC), and highlight the benefits of DcC over ScC. Therefore, we first provide a brief description of ScC algorithm in the following.

\subsection{Source-Centric Cooperation (ScC)}
\vspace{-5pt}
\begin{itemize}
 \item {\em Rate Control:} At every time slot $t$, the source node determines ${x}_k(t)$;
\begin{align} \label{eq:ScCrate_control}
\max_{\boldsymbol {{x}}} & \mbox{  }  [M U_{k}({x}_k(t)) -  {\mu}_{k}(t) {x}_{k}(t) ] \nonumber \\
\mbox{s.t. } &  {x}_k(t) \leq R_{k}^{max}
\end{align}
 \item {\em Cellular Link Scheduler:} At every time slot $t$, the source node determines ${x}_{k,k}(t)$ and ${x}_{n,k}(t)$;
\begin{align} \label{eq:ScCcellular_scheduling}
\max_{\boldsymbol {{x}}} & \mbox{  }   {\mu}_{k}(t) {x}_{k,k}(t) +  \sum_{n \in \Nset-\{k\}} ({\mu}_{k}(t) - {\nu}_{n,k}(t)){x}_{n,k}(t) \nonumber \\
  \mbox{s.t. } & \{{x}_{n,k}(t)\}_{\forall n \in \Nset} \in \Gamma_{\boldsymbol C^c(t)}.
  \end{align}
 \item {\em Cooperation \& Local-Area Link Scheduler for Unicast:} At time slot $t$, node $n$ determines the link rate ${h}_{n,k}(t)$;
\begin{align} \label{eq:ScClocal_area_scheduling_unicast}
 \max_{\boldsymbol {{h}}} & \mbox{  }  \sum_{k \in \Nset} \sum_{n \in \Nset-\{k\}} {\nu}_{n,k}(t) {h}_{n,k}(t)   \nonumber \\
\mbox{s.t.} \mbox{ }  & \{{h}_{n,k}(t)\}_{\forall k \in \Nset, n \in \Nset -\{k\}} \in \Gamma_{\boldsymbol C_{u}^{w}(t)}.
\end{align}
 \item {\em Cooperation \& Local-Area Link Scheduler for Broadcast:}
At time slot $t$, node $n$ determines the broadcast rate;
\begin{align} \label{eq:ScClocal_area_scheduling_broadcast}
 \max_{\boldsymbol {{f}}} & \mbox{  }  \sum_{k \in \Nset} \sum_{n \in \Nset-\{k\}} \sum_{\Jset in \Hset | k \in \Jset, n \notin \Jset}  {\nu}_{n,k}(t) {f}_{n,J}(t)   \nonumber \\
\mbox{s.t.} \mbox{ }  & \{{f}_{n,\Jset}(t)\}_{\forall n \in \Nset, \Jset \in \Hset | k \notin \Jset} \in \Gamma_{\boldsymbol C_{b}^{w}(t)}
\end{align} where ${h}_{n,k}(t) = \sum_{\Jset \in \Hset | k \in \Jset, n \notin \Jset} f_{n,\Jset}(t)$.
\end{itemize}

\subsection{Benefits of DcC over ScC}
In this section, we explain the benefits of DcC over ScC in terms of overhead, delay, and practical deployment.

{\em Overhead:} ScC determines ${x}_k(t)$, ${x}_{k,k}(t)$, and ${x}_{n,k}(t)$ at the source node according to Eqs.~(\ref{eq:ScCrate_control}), and (\ref{eq:ScCcellular_scheduling}). Therefore, the source node should know the queue sizes; ${\mu}_{k}(t)$, ${\nu}_{n,k}(t)$, and cellular downlink properties $\Gamma_{\boldsymbol C^c(t)}$. Although ${\mu}_{k}(t)$ is constructed at the source node, ${\nu}_{n,k}(t)$ is constructed at mobile devices, and the cellular downlink properties $\Gamma_{\boldsymbol C^c(t)}$ are usually measured by mobile devices. Therefore, ${\nu}_{n,k}(t)$ and $\Gamma_{\boldsymbol C^c(t)}$ should be carried to the source node from each mobile device over a cellular uplink. These control messages introduce $O(N)$ overhead over each cellular uplink.

On the other hand, in DcC, mobile devices construct all the real and virtual queues and make all decisions. \Eg mobile device $k$ determines and requests $x_{k}(t)$ and $\max_{n \in \Nset}\{x_{k,n}(t)\}$ amount of video packets from the source. These request messages introduce $O(1)$ overhead over each cellular uplink. Thus, DcC reduces the overhead from $O(N)$ to $O(1)$, which is significant considering the fact that cellular link capacities are limited as the demand for cellular links is already high and keeps increasing \cite{cisco_index}, \cite{ericsson_report}. Furthermore, since DcC introduces constant overhead over the cellular links, it provides scalability.

{\em Delay:} DcC improves packet delay over ScC thanks to employing virtual queues. Indeed, although the virtual queue sizes could be large in DcC, the real queue sizes could be significantly small as compared to the real queue sizes in ScC. Furthermore, the loss of control packets carrying queue size and cellular link quality information over cellular links increases real queue sizes in ScC. On the other hand, DcC makes all the decisions using local information in the mobile devices, so control packets are not carried over cellular links (only packet request messages are carried over the cellular links in DcC), so the loss of control packets does not affect DcC as much as ScC. The simulation results provided in the next section demonstrate the benefit of DcC in terms of delay as compared to ScC.

{\em Practical Deployment:} With the introduction of Dynamic Adaptive Streaming over HTTP (DASH) or MPEG-DASH \cite{dash}, there is an increasing interest to client-based video streaming applications, \eg Netflix uses DASH \cite{netflixdash}. According to DASH, the clients request video chunks at different rates using their connection level measurements. Our device-centric approach, since it operates at the client side, could be easily engaged with DASH to develop cooperative video streaming applications. Note that this could not be possible in ScC as it requires the video servers to be involved in the decision of which video chunks should be transmitted to the clients. We believe that our approach could be used to extend DASH for cooperative video streaming in mobile devices.

\subsection{Simulation Results}
In this section, we demonstrate the benefits of DcC over ScC in terms of overhead and delay through simulations. We consider a cooperative video streaming system and topology shown in Fig.~\ref{fig:forward_backward-system} for different number of users.

\begin{figure*}[t!]
\centering
\subfigure[DcC]{{\includegraphics[width=3.8cm]{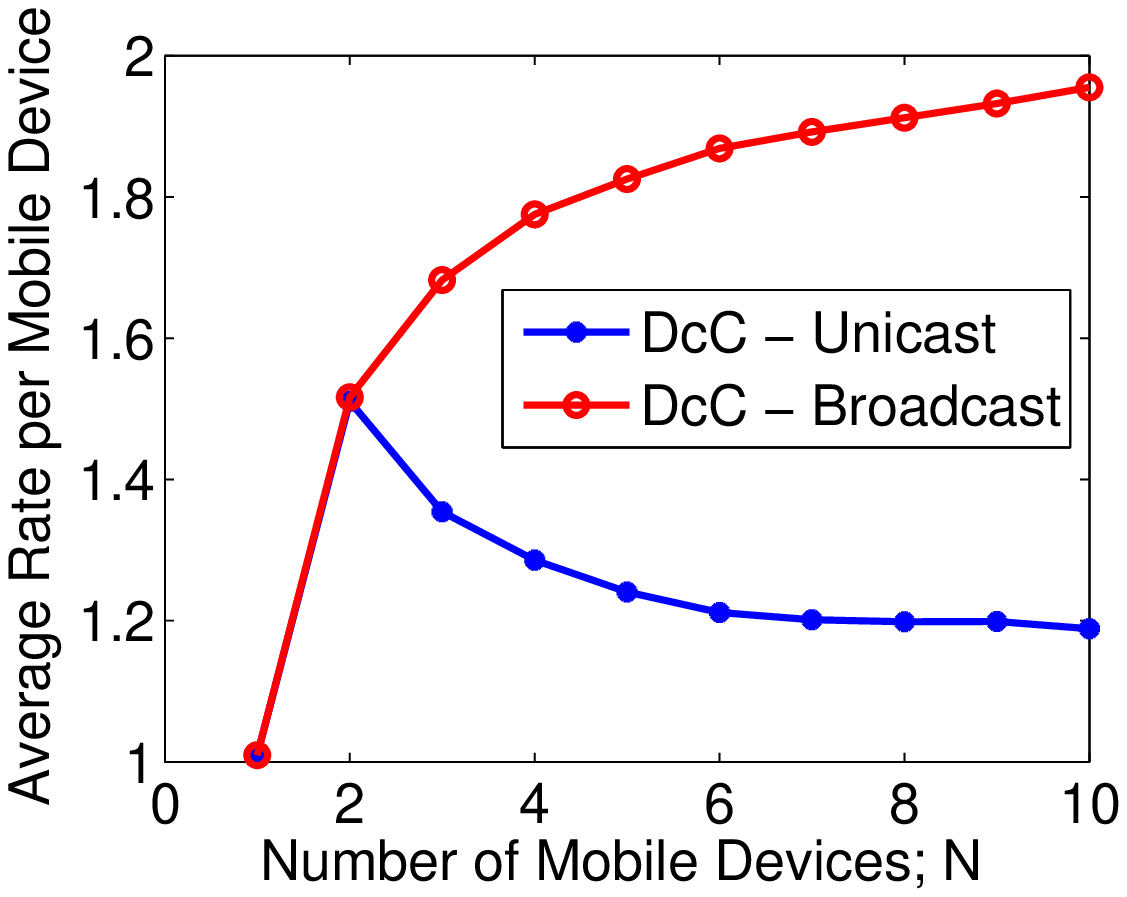}}}
\subfigure[ScC]{{\includegraphics[width=3.8cm]{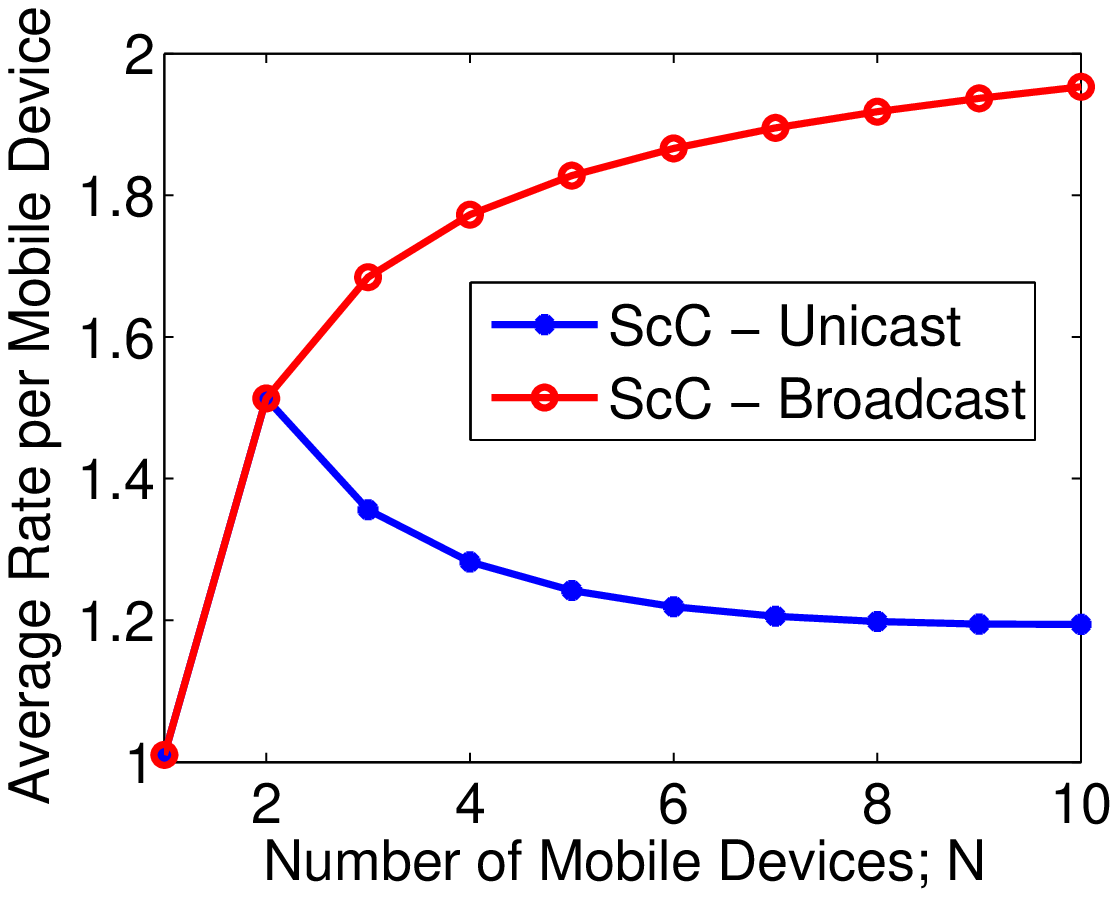}}}
\subfigure[Overhead] {{\includegraphics[width=3.8cm]{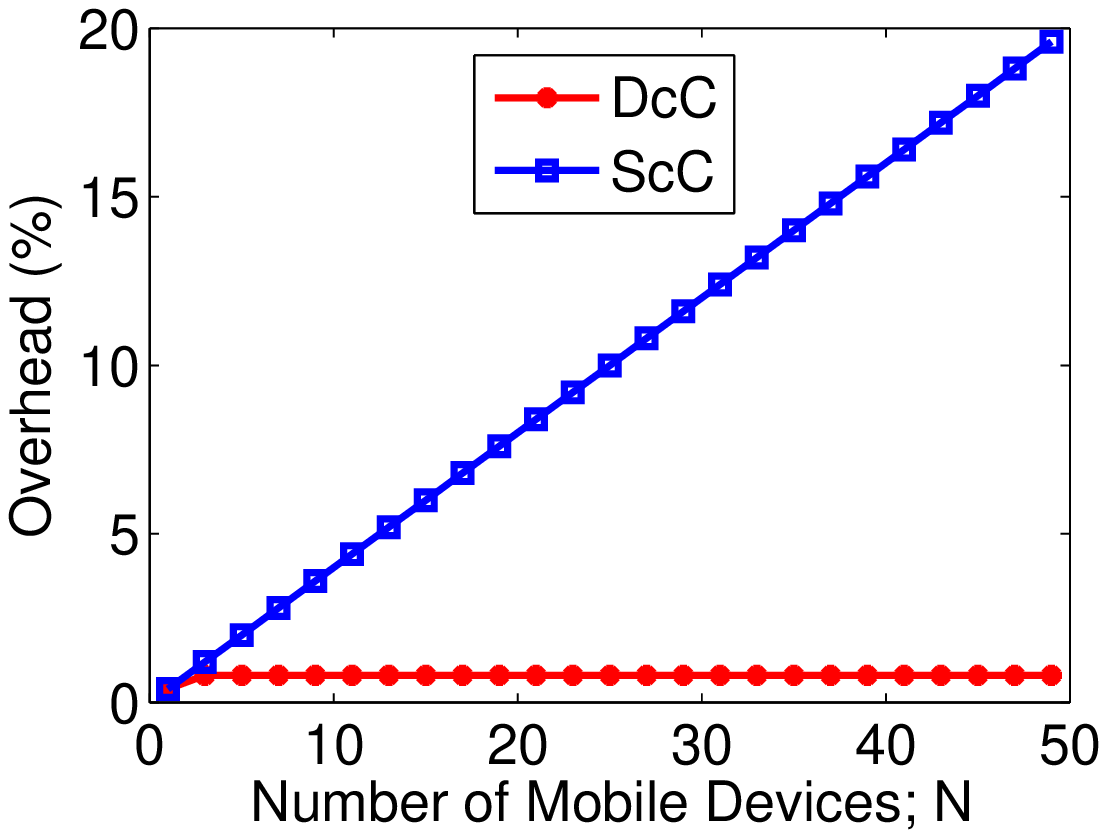}}}
\vspace{-5pt}
\caption{Average rate per mobile device in unicast and broadcast scenarios for (a) DcC and (b) ScC. (c) Percentage of overhead vs packet size.}
\label{fig:sims_DcC_ScC_Rate}
\end{figure*}

Fig.~\ref{fig:sims_DcC_ScC_Rate} presents the average rate per mobile device versus number of users for DcC and ScC. In this setup, the cellular and local area link rates are the same and 1 unit, and there is no loss over the links. As seen, in both DcC and ScC, broadcast improves over unicast as local area resources are used more efficiently. More importantly, DcC and ScC achieve the same rates for both unicast and broadcast, which is expected from Theorem~\ref{eec_theorem2}. Note that we do not take into account the effect of overhead in this simulation, \ie the length of control packets are zero bytes.

\begin{figure*}[t!]
\centering
\subfigure[ScC - $\mu_k(t)$]{{\includegraphics[width=3.8cm]{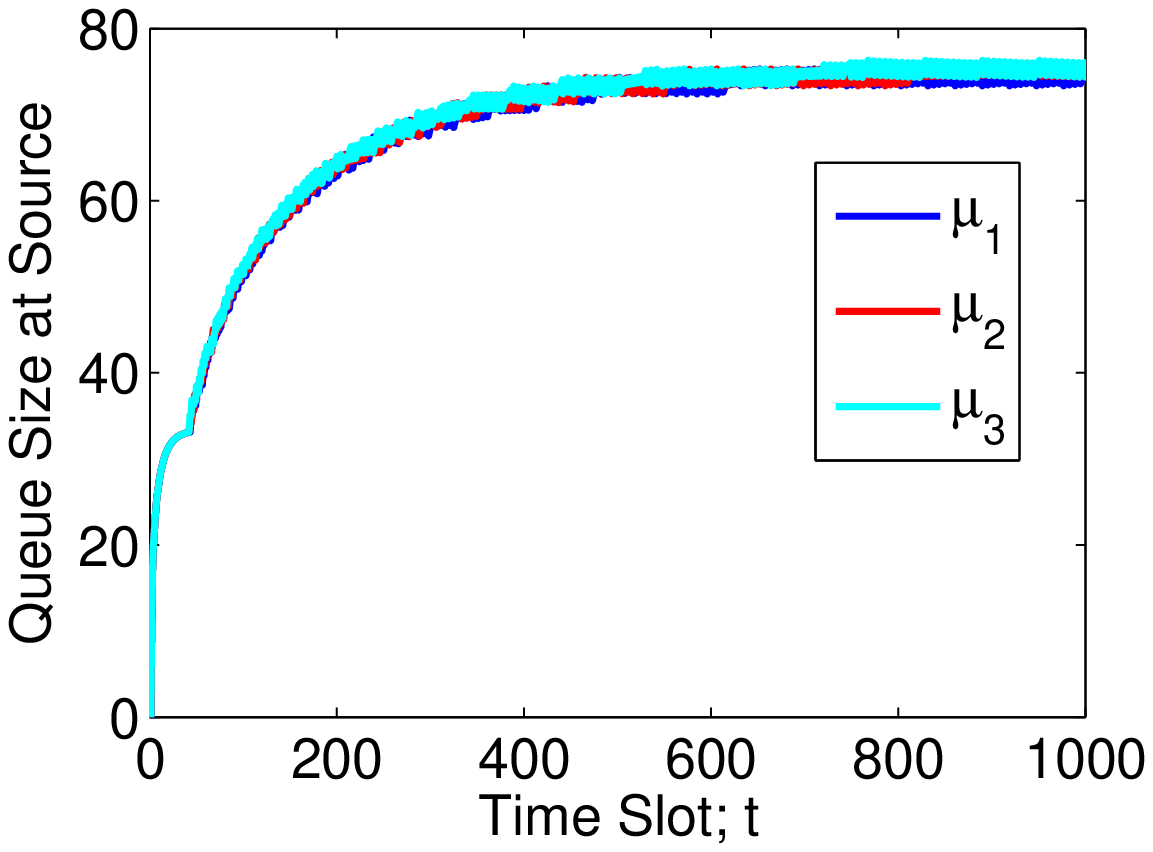}}}
\subfigure[ScC - $\nu_{n,k}(t)$]{{\includegraphics[width=3.8cm]{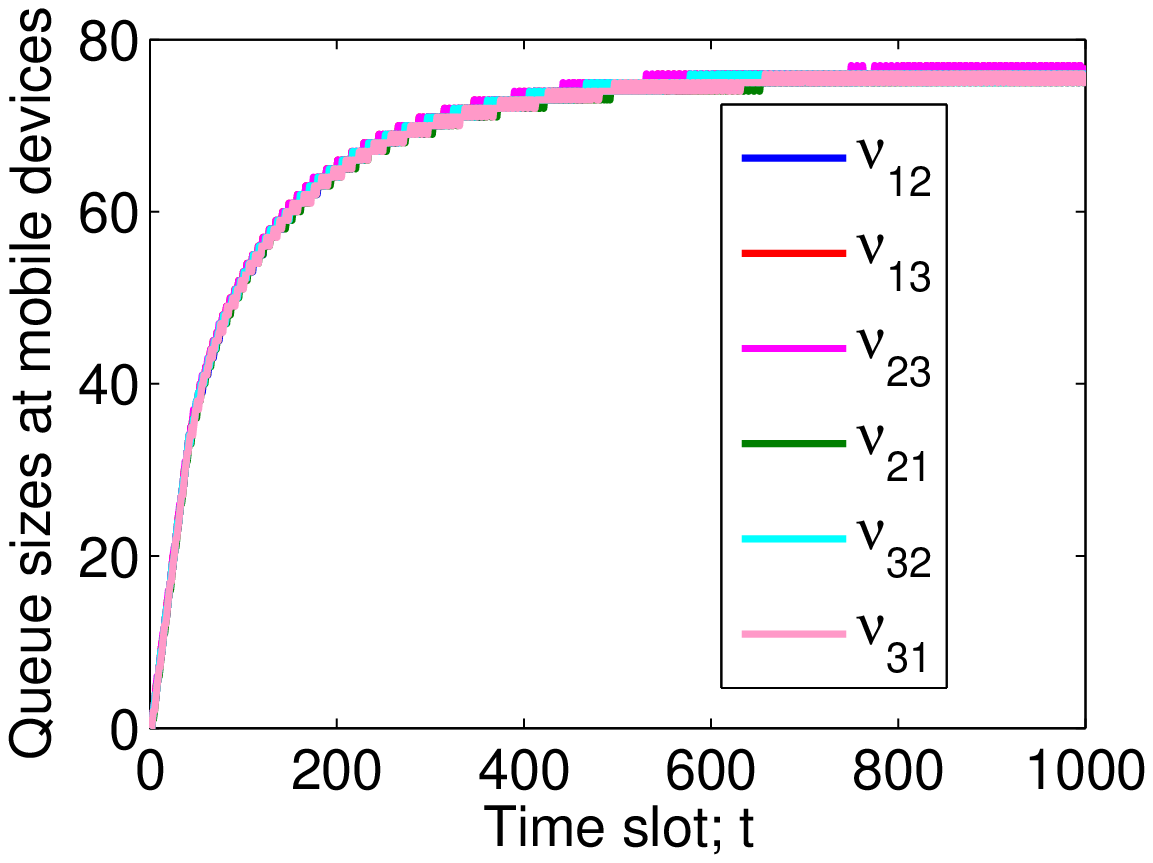}}}
\subfigure[DcC - Real Queues]{{\includegraphics[width=3.8cm]{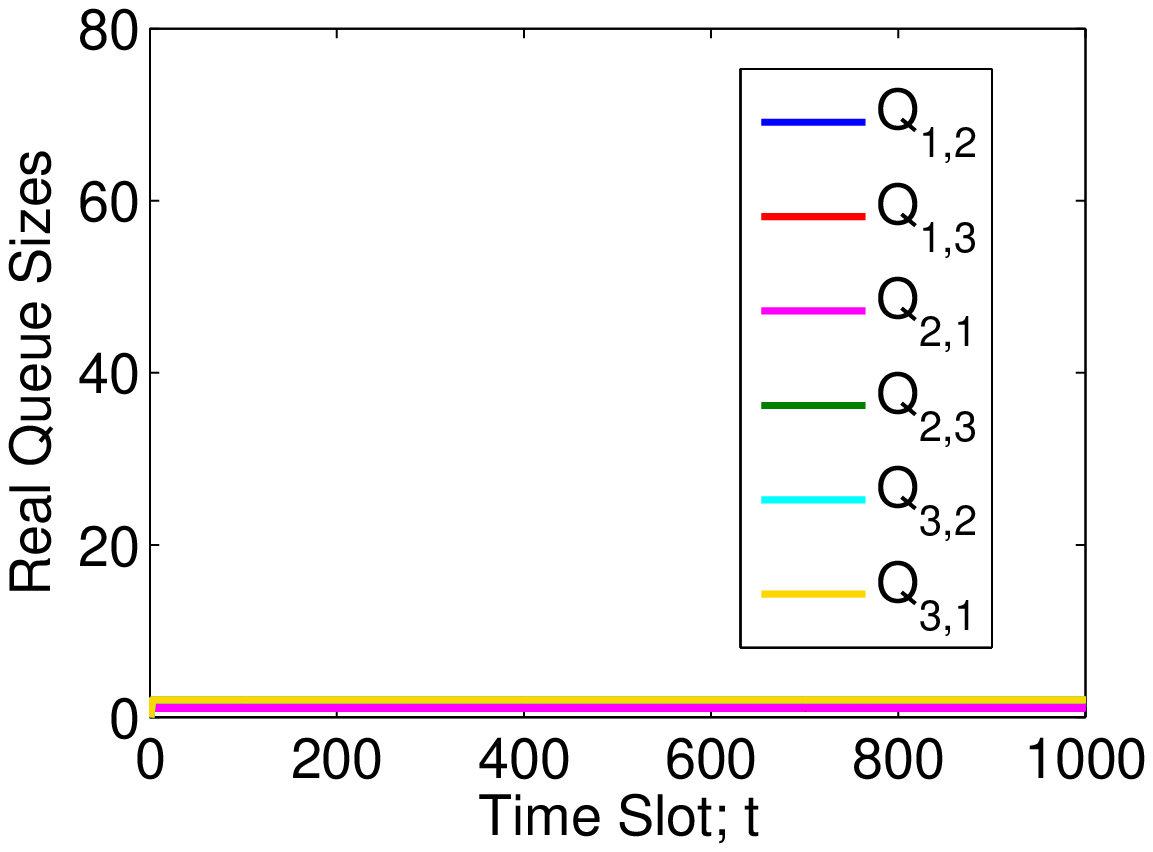}}}
\subfigure[DcC - Virtual Queues]{{\includegraphics[width=3.5cm]{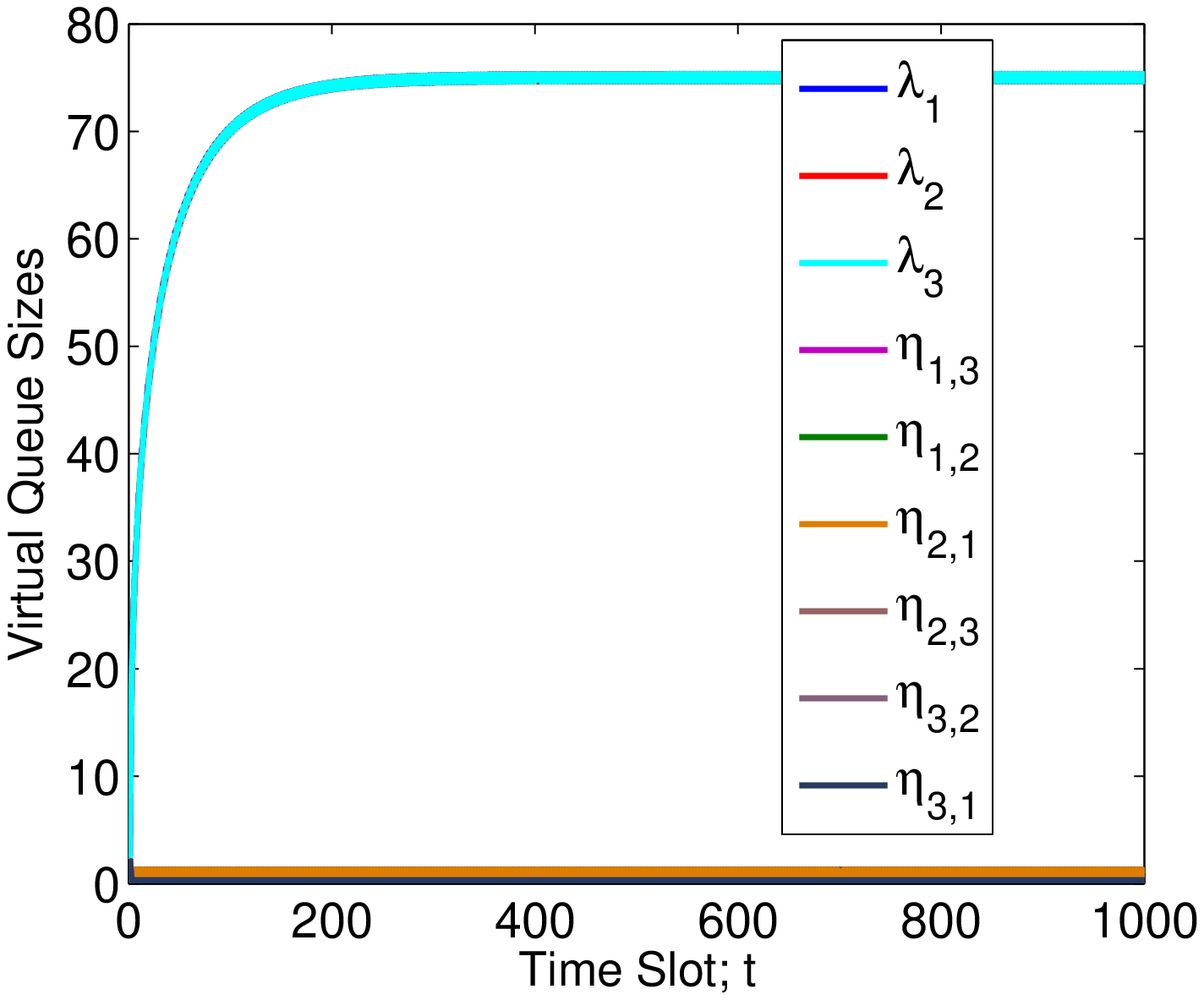}}}
\vspace{-5pt}
\caption{Queue Sizes. (a) ScC. Queue sizes at the source. (b) ScC. Queue sizes at the mobile devices. (c) DcC. Real queue sizes at the mobile devices. (d) Virtual queue sizes at the mobile devices.}
\vspace{-5pt}
\label{fig:sims_queueSizes}
\end{figure*}

Let us now consider overhead. We consider that queue size and channel state information are carried using 4 bytes from the mobile devices to the video servers in ScC, and the video rate request messages are carried from the mobile devices to the video servers using 4 bytes in DcC. The percentage of the overhead as compared to packet size, which we assume to be 1000 bytes is presented in Fig.~\ref{fig:sims_DcC_ScC_Rate}(c). The overhead of ScC is increasing with the increasing number of users, while the overhead does not change with the increasing number of users for DcC. For example, the overhead is almost 20\% when the number of mobile devices is 50. This means that 20\% of the cellular link capacities should be allocated to carry the control messages in ScC. On the other hand, the overhead of DcC is small for any number of mobile devices.

Fig.~\ref{fig:sims_queueSizes} presents queue size vs time for DcC and ScC. In this setup, both cellular and local area link rates are 1 units, and there is no loss over the links. As seen, the real queue sizes of ScC; \ie $\mu_{k}(t)$ and $\nu_{k}(t)$, could be very large, up to 75 packets. On the other hand, although virtual queue sizes could be also large in DcC, the real queue sizes; $Q_{n,k}(t)$ is very low. Thus, our scheme reduces queueing delay. 

\begin{figure}[t!]
\centering
\subfigure[DcC]{{\includegraphics[width=4.3cm]{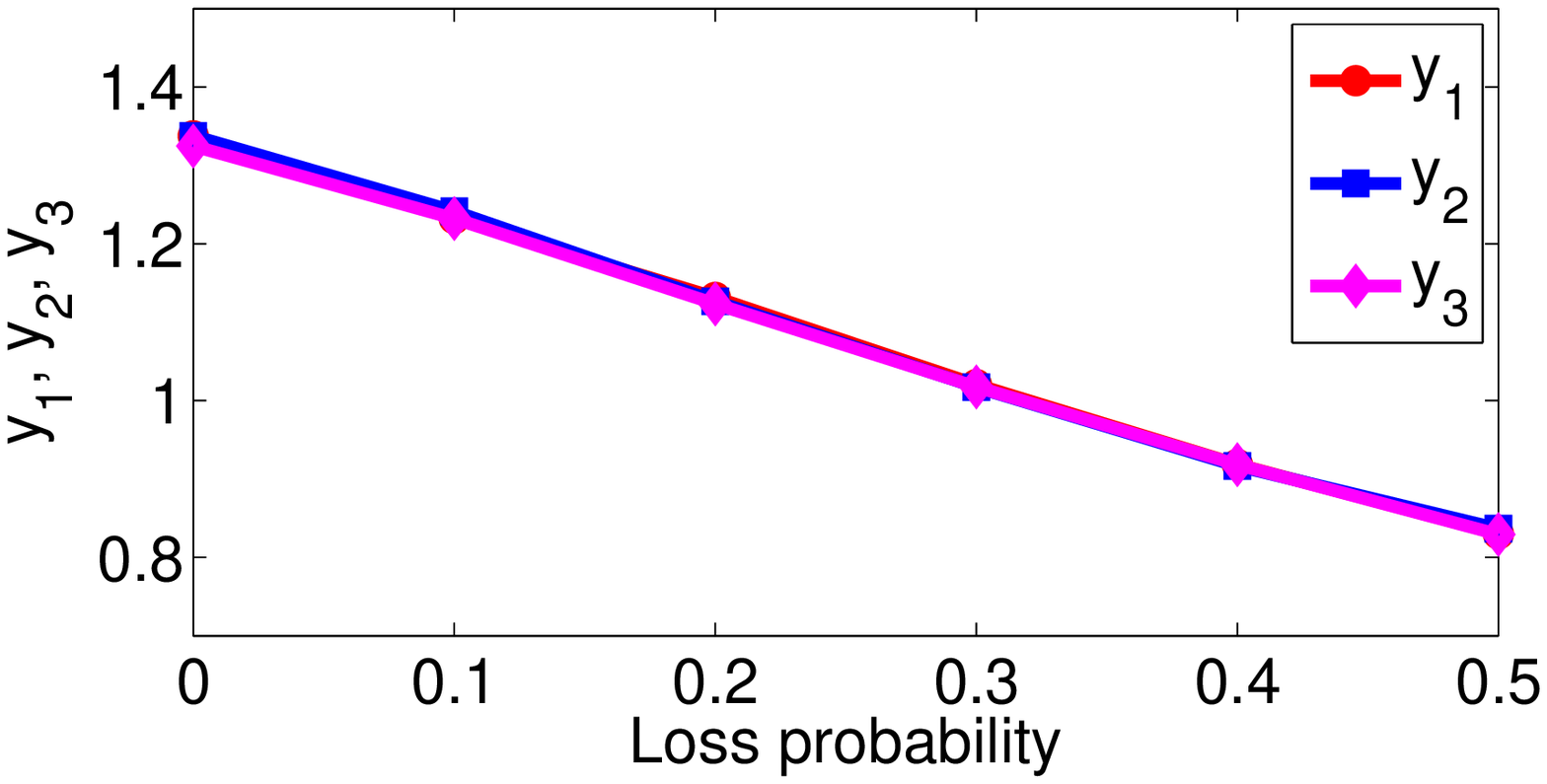}}}
\subfigure[ScC]{{\includegraphics[width=4.3cm]{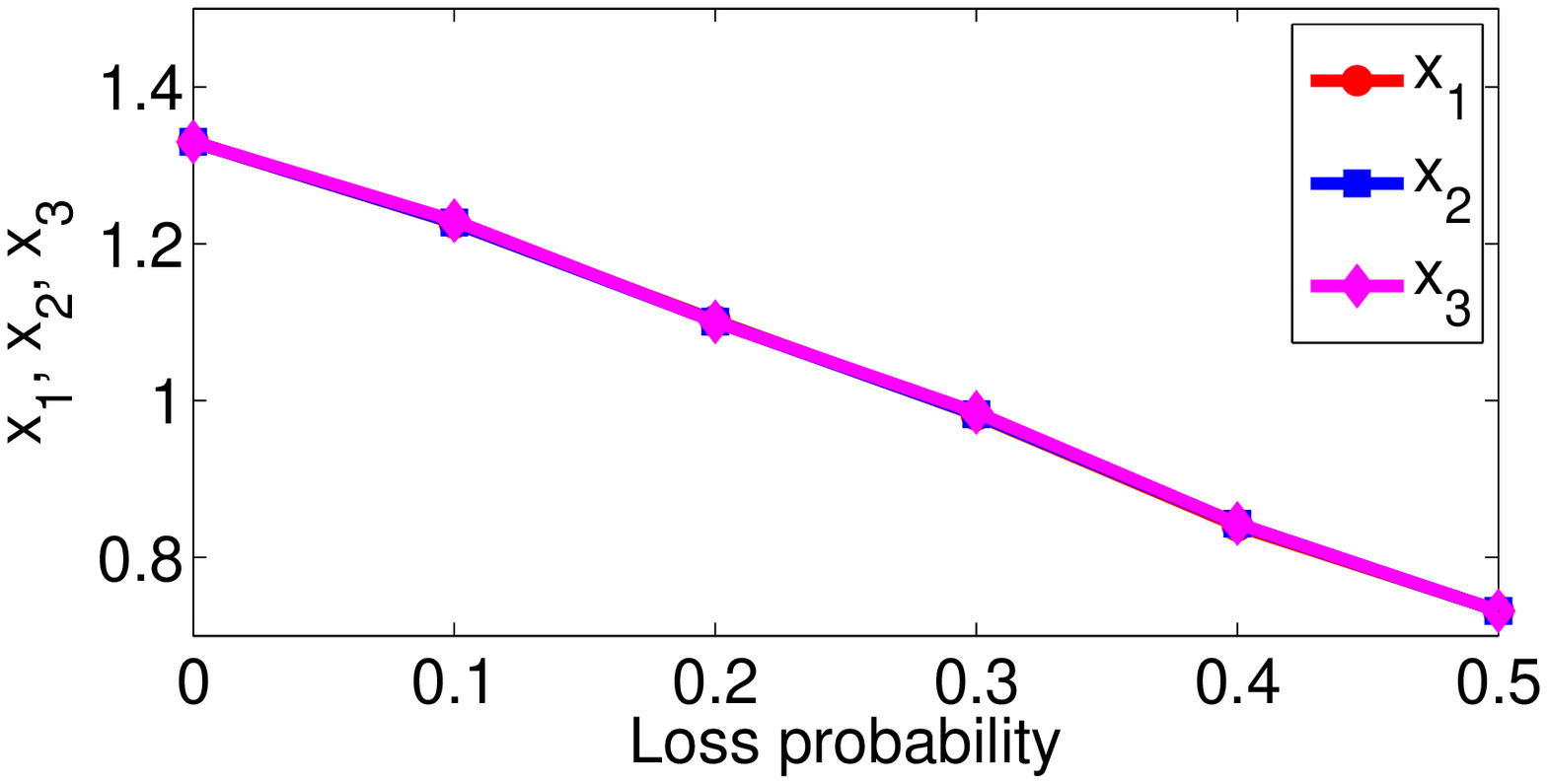}}}
\vspace{-5pt}
\caption{ Rate versus loss probability over the cellular links. (a) DcC. (b) ScC.}
\vspace{-5pt}
\label{fig:sims_rateVsCellularLoss}
\end{figure}

\begin{figure}[t!]
\centering
\subfigure[DcC]{{\includegraphics[width=4.3cm]{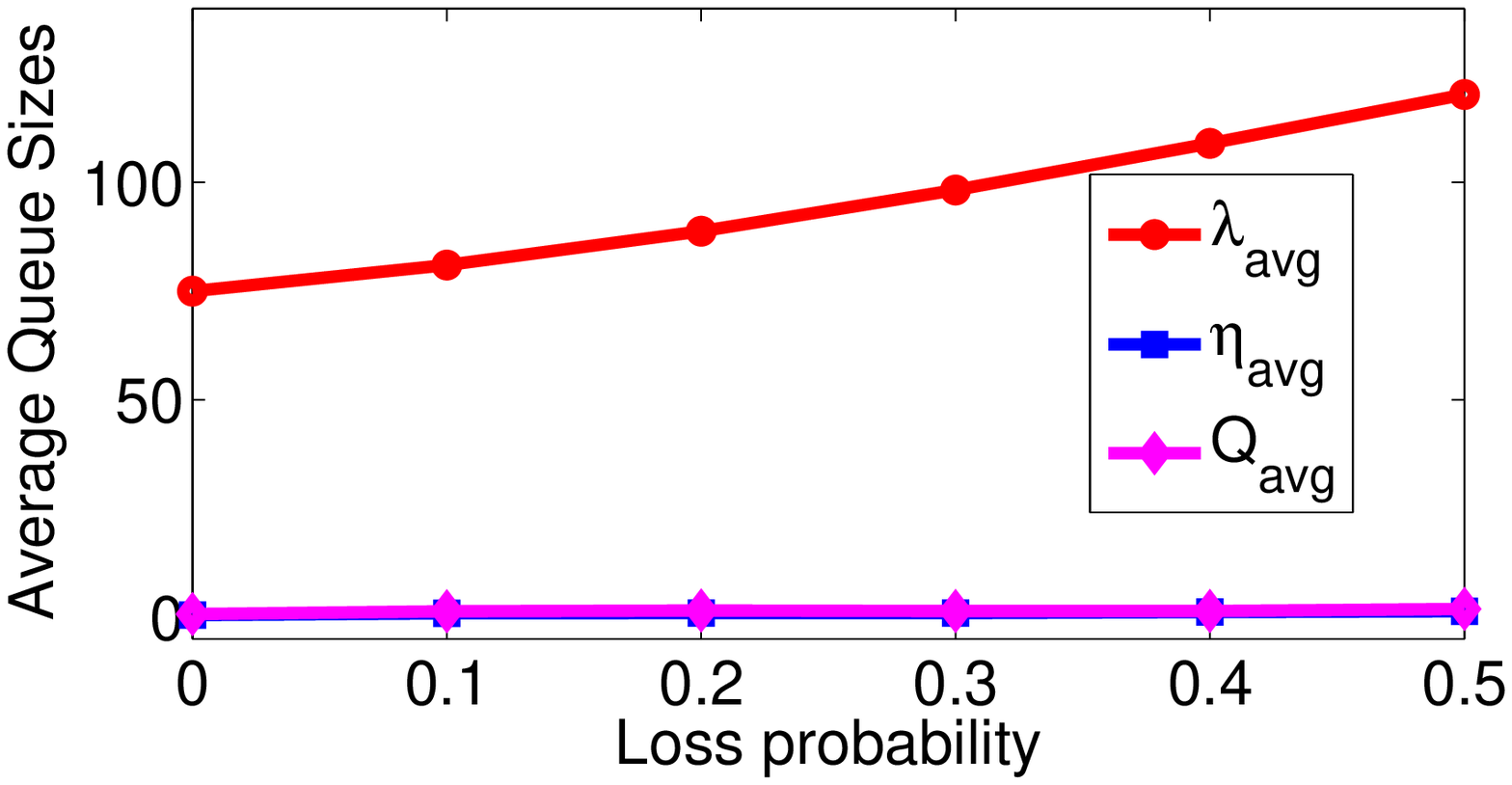}}}
\subfigure[ScC]{{\includegraphics[width=4.3cm]{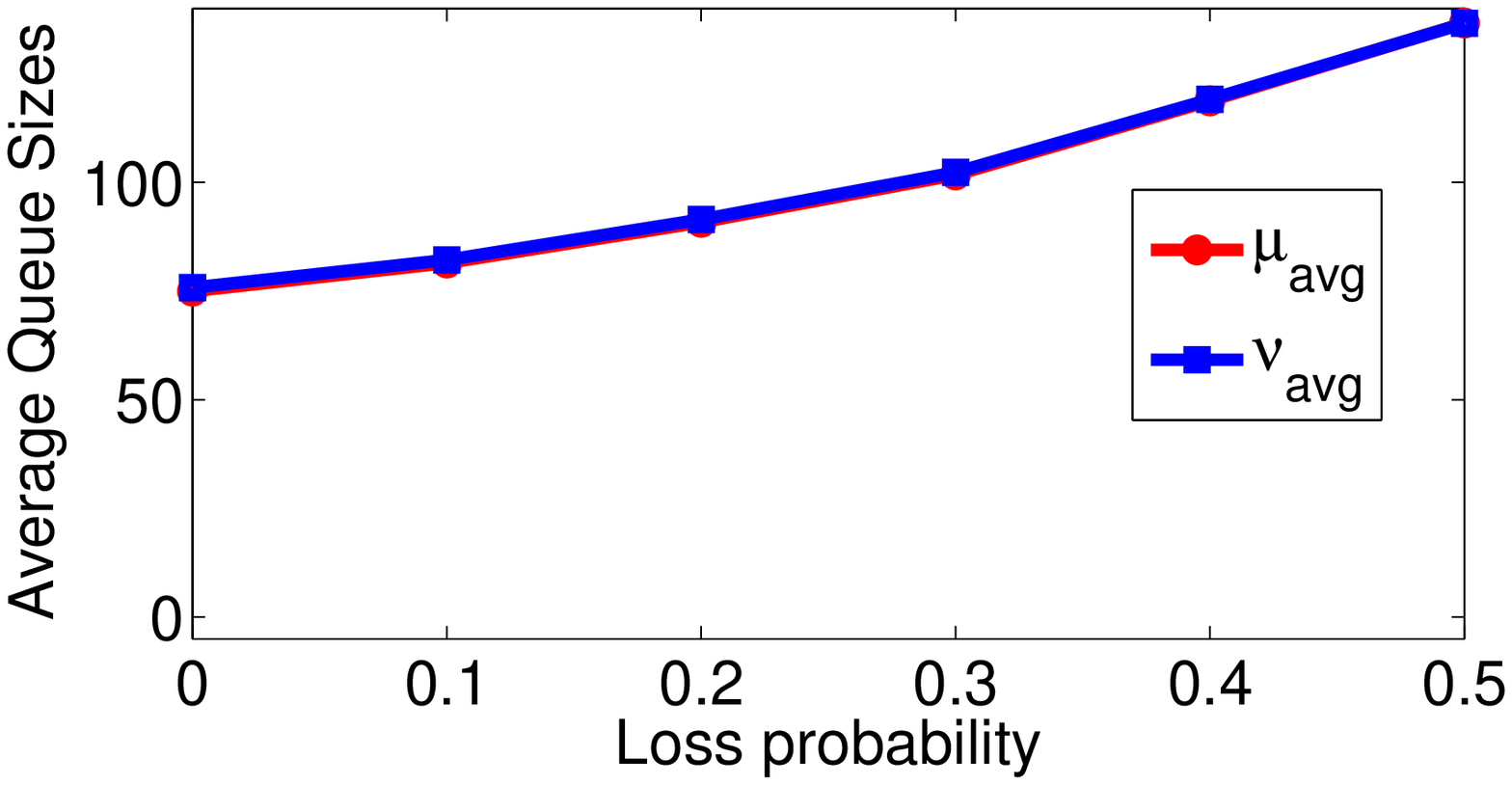}}}
\vspace{-5pt}
\caption{ Rate versus loss probability over the cellular links. (a) DcC. (b) ScC.}
\vspace{-5pt}
\label{fig:sims_avgQueueSizeVsCellularLoss}
\end{figure}

Fig.~\ref{fig:sims_rateVsCellularLoss} presents transmission rate towards each user versus the loss probability over the cellular links.  In this setup, both cellular and local area link rates are 1 units, and there is loss only over the cellular links, \ie there is no loss over the local-area links. As expected, in both DcC and ScC, flow rates decrease with increasing loss probability. However, DcC improves over ScC when the loss rate increases, because control packets are lost over the cellular links at high loss rates, and the source cannot make correct decisions in ScC. Fig.~\ref{fig:sims_avgQueueSizeVsCellularLoss} shows the average queue size versus the loss probability for the same setup. In particular, queue sizes are averaged over time and per-node queues. For example, $\lambda_{avg}$ is the average queue size of $\lambda_1$, $\lambda_2$, and $\lambda_3$ which are time averages of $\lambda_1(t)$, $\lambda_2(t)$, and $\lambda_3(t)$, respectively. As seen, although the virtual queue sizes increase in DcC with the increasing loss probability, the real queue size $Q_{avg}$ is very small and does not really increase with the increasing loss probability. On the other hand, the queue sizes in ScC, which are already very high as compared to DcC, increase significantly with increasing loss rate, which introduces significant delay.

\section{\label{sec:related} Related Work}
This work combines ideas from cooperation, network utility maximization, and stochastic network control.

When several users are interested in the same content, cooperative streaming is promising to improve throughput. For instance, \cite{micro1}, \cite{micro2}, \cite{micro3} consider a scenario in which device-to-device and cellular connections are used to disseminate the content, considering the social ties and geographical proximity for cooperation. Cooperation between mobile devices for content dissemination taking into account social ties, has been studied extensively  \cite{micro4,micro5}. Cooperative video streaming systems are implemented over mobile devices in \cite{micro6, micro7}. As compared previous work, the goal of this paper is to design device-centric cooperation scheme.

The NUM framework is promising to understand how different layers and/or algorithms, such as flow control, congestion control, and routing should be designed and optimized \cite{tutorial_doyle}, \cite{tutorial_lin}. We follow a similar approach, but we formulate the NUM framework considering the specific requirements such as device-centric design of the cooperative mobile devices.

The traditional source-centric, and backpressure-based stochastic network control algorithms have emerged from the pioneering work in \cite{tass1}, \cite{tass2}, which showed that in wireless networks where nodes route packets and make scheduling decisions based on queue backlog differences, one can stabilize queues for any feasible traffic. It has also been shown that backpressure can be combined with flow control to provide utility-optimal operation guarantee \cite{neelymoli}. Recently, receiver-based flow control scheme is developed for overloaded networks \cite{limod}. As compared to previous work, our scheme is designed for cooperative mobile devices, and it creates virtual flows and queues to move control functionality to mobile devices, and reduces the overhead over cellular links and delay, which was not the focus of the previous work.

\section{\label{sec:conclusion}Conclusion}
In this paper, we considered a cooperation scenario among mobile devices for video streaming. We developed a device-centric cooperation scheme; DcC. We showed that DcC reduces; (i) overhead; \ie the number of control packets that should be transmitted over cellular links, and (ii) the amount of delay that each packet experiences. Simulations demonstrate significant improvement in terms of overhead and delay.

\bibliographystyle{IEEEtran}

\begin{thebibliography}{}

\bibitem{cisco_index} Cisco Visual Networking Index: Global Mobile Data Traffic Forecast Update, 2010 - 2015.

\bibitem{ericsson_report} Ericsson Mobility Report, November 2013.



\bibitem{microcast} L.~Keller, A.~Le, B.~Cici, H.~Seferoglu, C.~Fragouli, A.~Markopoulou, ``MicroCast: Cooperative Video Streaming on Smartphones'' {\em in Proc. of ACM MobiSys}, Low Wood Bay, Lake District, UK, June 2012.

\bibitem{microcast_allerton} H.~Seferoglu, L.~Keller, B.~Cici, A.~Le, A.~Markopoulou,``Cooperative Video Streaming on Smartphones'' {\em in Proc. of Allerton}, 2011.

\bibitem{tutorial_lin} X. Lin, N. B. Schroff, R. Srikant, ``A tutorial on cross-layer optimization in wireless networks,'' {\em in IEEE JSAC}, vol. 24(8), Aug. 2006.

\bibitem{tutorial_doyle} M.~Chiang, S.~T.~Low, A.~R.~Calderbank, J.~C.~Doyle, ``Layering as optimization decomposition: a mathematical theory of network architectures,'' \emph{in Proceedings of the IEEE}, vol.~95(1), Jan. 2007.


\bibitem{thisTechRep} H.~Seferoglu, Y.~Xing, ``Device-Centric Cooperation in Mobile Networks,'' Tech. Report, available at {\em http://www.mit.edu/{\texttildelow}hseferog/}.

\bibitem{micro1} S.~Ioannidis, A.~Chaintreau, L.~Massoulie, ``Optimal and scalable distribution of content updates over a mobile social network,'' {\em in Proc. of INFOCOM}, Rio de Janeiro, Brazil, Apr. 2009.

\bibitem{micro2} B.~Han, P.~Hui, V.~A.~Kumar, M.~V.~Marathe, G.~Pei, A.~Srinivasan, ``Cellular traffic offloading through opportunistic communications: a case study,'' {\em in Proc. of ACM Workshop on Challenged Networks (CHANTS)}, Chicago, IL, Sept. 2010.

\bibitem{micro3} J.~Whitbeck, M.~Amorim, Y.~Lopez, J.~Leguay, V.~Conan, ``Relieving the wireless infrastructure: When opportunistic networks meet guaranteed delays,'' {\em in Proc. of IEEE WoWMoM}, Lucca, Italy, June 2011.

\bibitem{micro4} P.~Hui, J.~Crowcroft, E.~Yoneki, ``Bubble rap: social-based forwarding in delay tolerant networks,'' {\em in Proc. of ACM MobiHoc}, Hong Kong, May 2008.

\bibitem{micro5} C.~Boldrini, M.~Conti, A.~Passarella, ``Exploiting users' social relations to forward data in opportunistic networks: The HiBOp solution,'' {\em in Proc. of Pervasive and Mobile Computing}, Oct. 2008.


\bibitem{micro6} M.~Ramadan, L.~El Zein, Z.~Dawy, ``Implementation and evaluation of cooperative video streaming for mobile devices,'' {\em in Proc. of IEEE PIMRC}, Cannes, France, Sept. 2008.

\bibitem{micro7} S. Li and S. Chan, ``BOPPER: wireless video broadcasting with peer-to-peer error recovery,'' {\em in Proc. of IEEE ICME}, Beijing, China, July 2007.







\bibitem{gupta_interference_model} P.~Gupta, P.~R.~Kumar, ``The capacity of wireless networks,'' \emph{in IEEE Trans. on Information Theory}, vol.~34(5), 2000.

\bibitem{dash} ISO/IEC JTC1/SC29/WG11, ``Information technology - Dynamic adaptive streaming over HTTP (DASH) -- Part 1: Media presentation description and segment formats'', ISO/IEC 23009-1:2012, 2012.

\bibitem{netflixdash} http://techblog.netflix.com/2010/12/html5-and-video-streaming.html

\bibitem{tass1} L. Tassiulas, A. Ephremides, ``Stability properties of constrained queueing systems and scheduling policies for maximum throughput in mul- tihop radio networks,'' {\em in IEEE Trans. on Automatic Control}, vol. 37(12), Dec. 1992.

\bibitem{tass2} L. Tassiulas and A. Ephremides, ``Dynamic server allocation to parallel queues with randomly varying connectivity,'' {\em in IEEE Trans. on Information Theory}, vol. 39(2), March 1993.

\bibitem{neelymoli} M. J. Neely, E. Modiano, and C. Li, ``Fairness and optimal stochastic control for heterogeneous networks,'' {\em in IEEE/ACM Trans. on Networking}, vol. 16(2), April 2008.

\bibitem{limod} C.-P.~Li and E.~Modiano, ``Receiver-Based Flow Control for Networks in Overload,'' {\em in IEEE/ACM Transactions on Networking}, 2014.

\bibitem{neelybook} M.~J.~Neely, ``Stochastic network optimization with application to communication and queueing systems,'' Morgan \& Claypool, 2010.


\end{thebibliography}


\end{document}